\documentclass[twoside,twocolumn,a4paper]{article}
\usepackage{times}
\usepackage{EAAproceedings}

\begin{document}

% ****************  Begin of manuscript **********************
\Englishtitle{Global and local synthetic descriptions of the piano soundboard}

% Short title:
\Kolumnentitel{Synthetic descriptions of the piano soundboard}


% Authors' names, affiliations and addresses
% optionally grouped if with different institutions.
% First group:
\AuthorsI{Kerem Ege}
\AddressI{Laboratoire Vibrations Acoustique - INSA Lyon, 25 bis avenue Jean Capelle, F-69621,
Villeurbanne Cedex, France}
%Second group: (activate if applicable)
\AuthorsII{Xavier Boutillon}
\AddressII{Laboratory for the Mechanics of Solids, École polytechnique, F-91128 Palaiseau Cedex, France}
%Department of Proceedings, 
%University of Bruckgrad, 
%Bruckgrad, Country2.}
% Third group (activate if applicable)
% \AuthorsIII{}
% \AddressIII{}

% English Abstract:
\Englishabstract{Up to around 1.1 kHz, the soundboard of the piano behaves like a homogeneous %isotropic
plate whereas upper in frequency, it can be described as a set of waveguides defined by the ribs. In consequence:\linebreak
a) The acoustical coincidence phenomenon is deeply modified in comparison with that occurring in homogeneous plates since the dispersion curve of a waveguide can present none, one, or two coincidence frequencies. This may result in a nonuniformity of the soundboard radiation in the treble range, corresponding to the so-called \emph{killer} octave, where a good sustain is difficult to obtain.\linebreak
b) The mobility (mechanical admittance) in the direction normal to the soundboard can be synthesised with only a small number of parameters. It compares well with published measurements (Giordano, JASA, 1998), in particular the step-like falloff of the local impedance due to the localisation of the waves between ribs.\linebreak
c) The synthesised mobility has the same features as those which can be derived independantly, according to Skudrzyk (JASA, 1980) and Langley (JSV, 1994). \linebreak
This approach avoids the detailed description of the soundboard, based on a very large number of parameters. It can be used to predict global changes of the driving point mobility, and possibly of the sound radiation in the treble range, resulting from structural modifications\footnote{Most of this work has been done as %XXXKE ("has been as", ou "has been done as" ?)
a doctoral thesis by the first author at the Laboratory for the Mechanics of Solids. Part of it was presented at the 20$^\text{th}$ International Symposium on Music Acoustics, held at Sydney and Katoomba, August 2010, and reported in the short communication~\cite{EGE2010_2}.}.}
\ScientificPaper

%%%%%%%%%%%%%%%%%%%%%%%%%%%%%%%%%%%%
\section{Introduction}
Like for any extended continuous linear structure, the dynamics of the piano soundboard can be described by a superposition of modes. This description may be used up to a frequency-domain where the response of the structure becomes more or less flat, that is where modes are not distinguishable one from each other due to the overlapping character of their individual responses. As frequency increases, the validity of the modal description becomes more and more sensitive to the details of the local description of the structure. This paper deals with the frequency-domain up to 10kHz. Since the piano soundboard has a modal density of roughly 0.02, about 200 modes are involved in the description... %XXXKE (Dire que cela fait plusieurs milliers de degrés de liberté ?)
Since neither the modal analysis nor the musical significance give cues that would help to sort out the huge number of modal parameters or establish a hiererarchy between them, the modal description, as such, is of little help for dealing with practical questions. However, the piano soundboard can be described, as a structure, with far less parameters: elastic constants of the materials, geometry. Given an appropriate modelling, describing the main features of the dynamics must therefore be possible with far less parameters than the overall number of degrees of freedom.

The piano soundboard is essentially an interface between the strings and the acoustical field. We emphasise here the string viewpoint and give less attention to the acoustical field. We propose dynamical descriptors that are \emph{complete} but do not deal with the \emph{details} of the dynamics: all global features are taken into account so that the dynamics can be solved at a given point (mechanical mobility of the soundboard, as seen by the string) but the solution does not take into account precise details of the string location or of the frequency response, for example. To this end, we have investigated the modal behaviour of an upright piano soundboard (see Figure~\ref{fig:table_exp}) by means of a recently published high-resolution modal analysis technique~\cite{EGE2009}. The frequency evolution of the modal density of the piano soundboard reveals two well-separated vibratory regimes of the structure described in section~\ref{sec:vibregimes}. Consequences in terms of radiation of the soundboard are given. Based on the model corresponding to the findings of the modal analysis, the mobility (or mechanical admittance) at any point of the soundboard can be synthesised (section~\ref{sec:mobility}), in the spirit given in the introduction. It is compared to published measurements \emph{far from} the bridge and \emph{at} the bridge, and also to global approaches of the mobility of mechanical structures by Skudrzyk and Langley (section~\ref{sec:global}).
\Figure{fig:table_exp}{Rear view of the upright piano studied with the wood grain direction added in green and the two bridges (on the other side of the soundboard) added in red.}
  {50}{  \put(6,1)  {  \includegraphics[width=65mm]{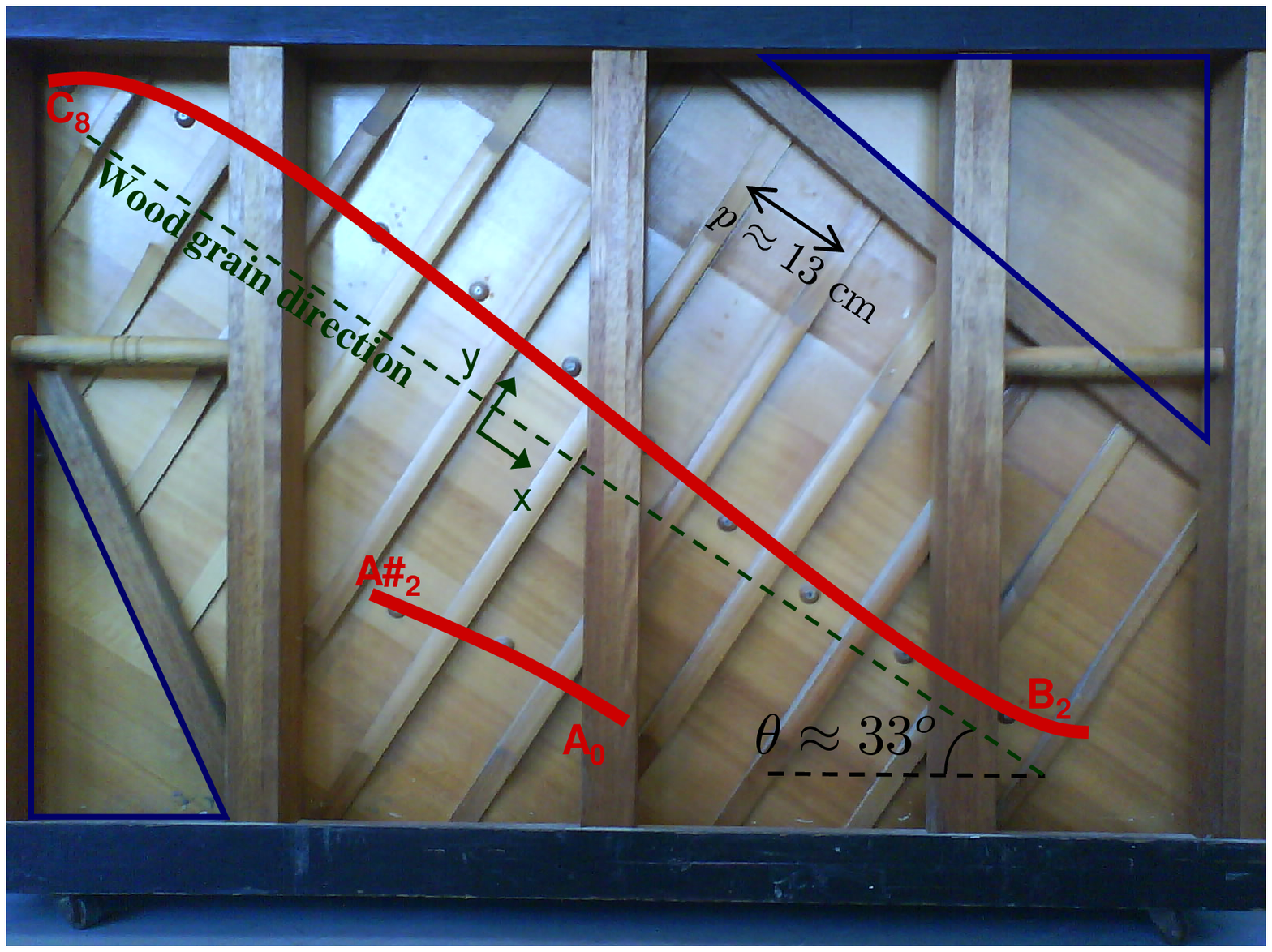}}}

\section{Two vibratory regimes}\label{sec:vibregimes}
\subsection{Modal density}
\label{sec:ModDens}
The modal density $n(f)$ is a global descriptor of the vibratory behaviour of the soundboard in the mid-frequency domain. Based on measured modal frequencies (\cite{EGE2009_2}), the modal density has been estimated as the moving average of the modal spacing (six successive modes retained for each estimation). The frequency evolution of $n(f)$ (Fig.~\ref{fig:modaldensity}) at four points of measurements (see Fig.~\ref{fig:table_exp} for the exact locations) reveals two distinct vibratory regimes of the structure.
\begin{enumerate}
\item{Below $1.1$~kHz, the four experimental curves are almost similar. The modal density increases slowly and tends towards a constant value of $\approx0.06~\text{modes}~\text{Hz}^{-1}$ independently of the zones of the board where the measurement is done. This means that the corresponding modes extend over the whole soundboard: the ribbed board behaves as a homogeneous %isotropic 
plate. The rise in $n$ in the lowest frequency range is characteristic of \emph{constrained} boundary conditions. The theoretical asymptotic modal density of a isotropic homogeneous plate is:
\begin{equation}
n_\infty(f)=\dfrac{S}{2}\sqrt{\dfrac{\rho\, h}{D}}
\label{eq:ModalDensAsIso}
\end{equation}
where $S$ and $h$ are respectively the plate area and thickness, $\rho$ is the material density, and $D=\dfrac{Eh^3}{12(1-\nu^2)}$ with $E$ and $\nu$, the Young's modulus and Poisson ratio. The ribbed zone was replaced by an isotropic thin plate with surface and surfacic area %(XXXKE pourquoi préciser surface and surfacic area?) Pour avoir séparément S et h XXXKE Ok !
equal to those of the corresponding part of the soundboard. The rigidity of the homogenised plate $D$ was somewhat arbitrarily taken so that its \emph{dynamical rigidity} $D/(\rho\,h)$ equals that of the orthotropic spruce plate in the direction of the grain: $D_{\text{hom}}=D_{x,\text{wood}}=\dfrac{E_x h^3}{12(1-\nu_{xy}\nu_{yx})}$. %XXX compléter les indices de nu XXXKE c'est fait
The mechanical characteristics of the strips of wood are derived from measurements made by Berthaut~\cite{BER2003} on spruce species selected for piano soundboards}.
\item{For frequencies above $1.1$~kHz, $n(f)$ decreases significantly and is not exactly the same at each location of the soundboard. The interpretation is that ribs confine wave propagation: the soundboard behaves as a set of structural wave-guides. The lowest frequency for which this phenomenon occurs is such that the inter-rib spacing $p$ corresponds to a half-wavelength: $k_{x}=\pi/p$ (see Fig.~\ref{fig:table_exp} for the directions of the $x$- and $y$-axes), corresponding roughly to 1.1 kHz in the soundboard case. The modal density of such a waveguide can be easily calculated. The red continuous curve in Fig.~\ref{fig:modaldensity} corresponds to three times the theoretical modal density of one waveguide and corresponds closely to the measured modal densities. The interpretation is that the motion is not strictly confined to the inter-rib region where it has been generated but extends over the two adjacent regions \emph{with the same wavenumber} $k_{x}=\pi/p$. Thus, a given point of the soundboard "sees" three times more modes than there are in a single wave-guide and the modal density is also to be multiplied by 3. Modes can be organised in families, corresponding to the successive transverse modes (like in pipes) whose wavenumbers are $k_{x,m}=m\pi/p$ with $m\in\mathbb{N}^*$. The asymptotic modal density of any family (\emph{i.e.} restricted to one given transverse mode) is that of a beam of length $L_y$ ($\omega^{-1/2}$ dependency):
\begin{equation}
n(f)\:\underset{f\longrightarrow +\infty}{\rightarrow}\:\cfrac{L_y}{\sqrt{2\pi}\sqrt{f}}\,\left(\cfrac{\rho\,h}{D_y}\right)^{1/4}
\label{eq:ModalDenWGLim}
\end{equation}
As frequency increases, successive families enter into the modal population of the guide, multiplying step by step the modal density given by Eq.~\ref{eq:ModalDenWGLim} by 2, 3, etc., as it will be seen further. The mechanical characteristics of the waveguide are the one of the interrib orthotropic spruce plate.
}
\end{enumerate}
\Figure{fig:modaldensity}{Modal densities of the upright piano sondboard studied. %XXXKE modify caption ?
({\color[rgb]{0,0,1}\tiny{$\bullet$}}),~{\color[rgb]{1,0,0}$\blacktriangle$},~$\blacktriangledown$),~({\color[rgb]{0,0.5,0}\scriptsize{$\ast$}})~: measurements at four points of the soundboard (different zones). {\color[rgb]{1,0,0}\raisebox{.15\baselineskip}{\rule{1.5mm}{1pt}\ \rule{0.2mm}{1pt}\ \rule{1.5mm}{1pt}}}~: theoretical modal density of the homogeneous equivalent clamped plate (see \S~\ref{sec:ModDens}). 
{\color[rgb]{1,0,0}\raisebox{.15\baselineskip}{\rule{4.5mm}{1pt}}}~: three times the theoretical modal density of one waveguide for the first transverse mode (1,$n$).}
  {50}{  \put(2,2)  {  \includegraphics[width=77mm]{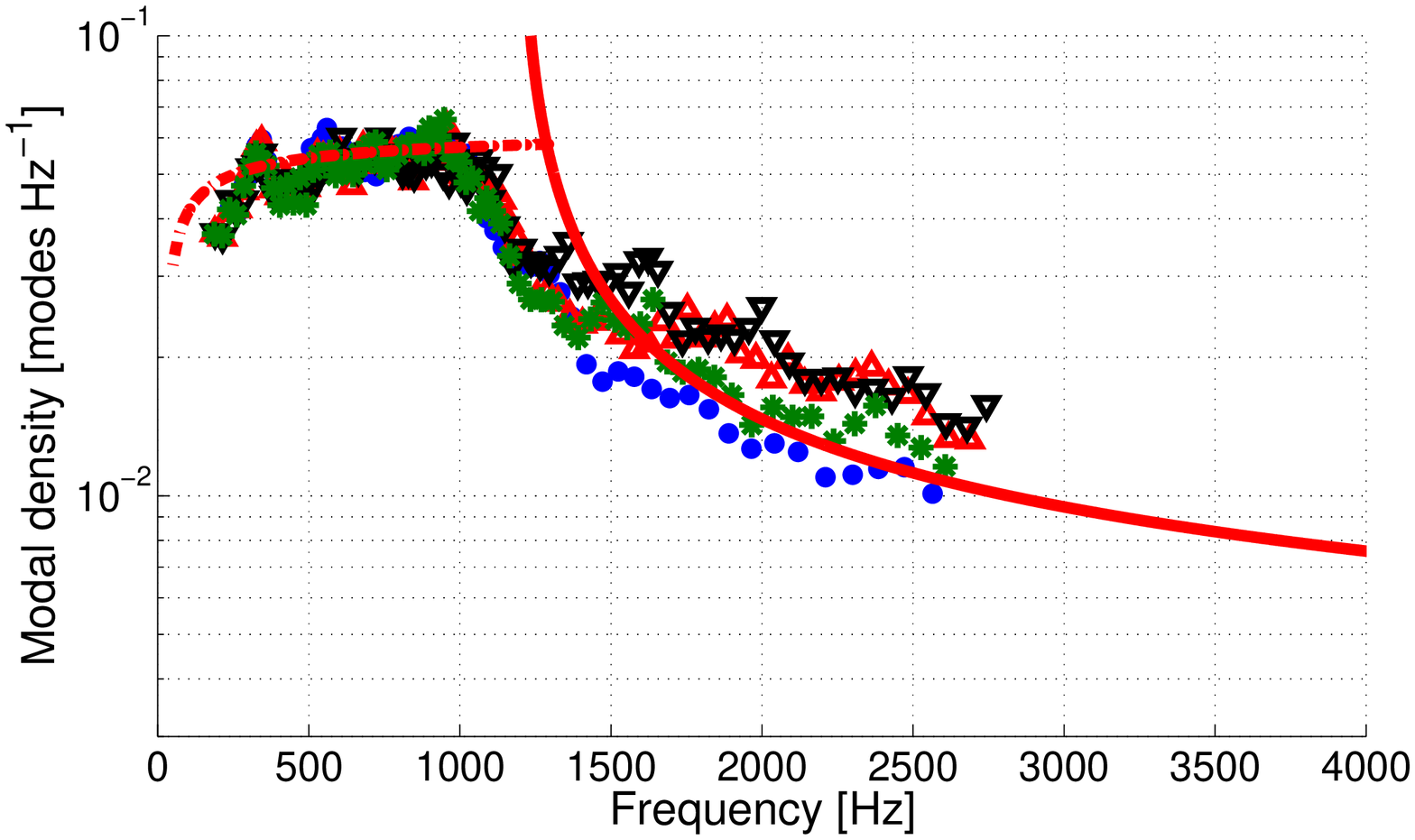}}}

\subsection{Nonuniformity of the radiation}
The structural wave-guide phenomenon deeply modifies the acoustical coincidence phenomenon in comparison to what occurs with a homogeneous plate. A plate radiates efficiently when the structural wavelength is larger than that in air (supersonic structural waves). For a thin isotropic plate, this occurs above a so-called coincidence frequency given by the intersection of the dispersion curves in the plate (blue plain and dotted line in Fig.~\ref{fig:dispers_guide}) and in air (dah-dot line in the same figure). Musical consequence: for a note with a fundamental below 1.1 kHz, the lowest partials radiate less efficiently (and thus, decrease in time less rapidly) that the upper ones.

The dispersion curve of a structural wave-guide is different: it always starts with \emph{supersonic} structural waves (efficient acoustical radiation) and may present two, one, or even no coincidence frequencies, depending on the value of the wave width $p$. Accordingly, there will be two, one, or no change in the radiation efficiency. This creates a nonuniformity in the radiation of the soundboard in the treble range of the instrument compared to the low-range which may explain the well-known difference in timbre. The timbre is influenced by the relative level of the partials and also by their relative time-decays. For example, for the key $\mathbf{\textbf{D}\sharp_6}$ having a fundamental frequency around 1245~Hz, the sound level \emph{and} the damping factor of the fundamental may be higher (due to the acoustical radiation: supersonic structural waves) than those of the next two partials (subsonic structural waves).

It may be possible to establish a connection between the frequency range where the radiation pattern of the piano is ruled by the wave-guide phenomenon (around 1.1 kHz) and the so-called \emph{killer} octave mentionned by some manufacturers. The transition between the two vibratory regimes of the soundboard and the induced nonuniformity of the acoustical radiation may explain why the sustain is so difficult to obtain around the fifth to sixth octave~(see for example comments of the Fandrich Piano Company's piano maker~\cite{FAN1995}).
\Figure{fig:dispers_guide}{Relations of dispersion for flexural waves in the homogeneous isotropic plate ({\color[rgb]{0,0,1}{---}}), in the air (--~$\cdot$~--) and for the two first modes ({\color[rgb]{0,0.5,0}{--~--}}) of the waveguide between the second and third ribs. $\bullet$ and $\circ$: discrete values corresponding to the $(1,n)$- and $(2,n)$-modes of the wave-guide, respectively. {\color[rgb]{1,0,0}$\diamond$}: partials of the~$\mathbf{\textbf{D}\sharp_6}$ strings. See also the range of the so-called \emph{killer} octave \textbf{A$_5$}-\textbf{A$_6$} ([880-1760]~\text{Hz}).}
  {43}{  \put(1,1)  {  \includegraphics[width=77mm]{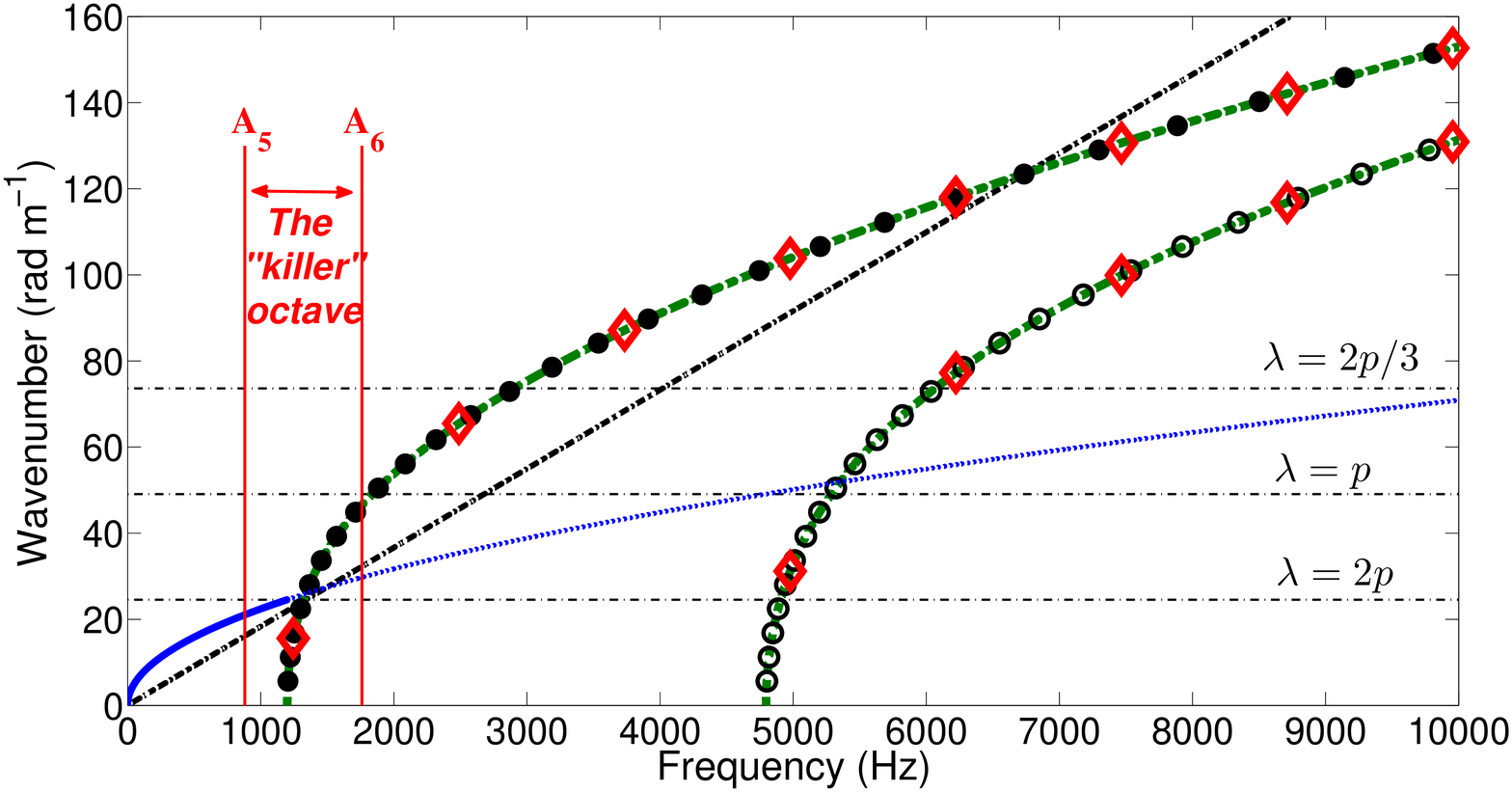}}}

\section{Synthesised mechanical mobility of the soundboard}

\label{sec:mobility}

The purpose here is to give an expression of the piano soundboard mechanical mobility (in the direction normal to the soundboard) depending on a small number of parameters and valid up to several kHz. 

\subsection{Analytical expression: sum of the modal contributions}\label{sec:sum}
The driving point mobility $Y_\text{A}$ at a point $(x_\text{A},y_\text{A})$ of a weakly dissipative vibrating system can be expressed as the sum of the mobility of single-degree of freedom linear damped oscillators:
\begin{equation}\label{eq:drive_admitt_amort}
Y_\text{A}(\omega)=\cfrac{V_\text{A}(\omega)}{F_\text{A}(\omega)}=i\omega\,\mathop{\sum}_{\nu=1}^{+\infty}\cfrac{\Phi_\nu^2(x_\text{A},y_\text{A})}{m_\nu\,(\omega_\nu^2+i\eta_\nu\omega_\nu\omega-\omega^2)}
\end{equation}
where $m_\nu$ is the modal mass, $\eta_\nu$ is the modal loss factor, $\omega_\nu$ the modal angular frequency and  $\Phi_\nu$ the modal shape of the mode $\nu$.

\subsection{Synthesis in the low-frequency range}
\label{sec:MobLowFreq}
In the low-frequency range, the vibration extends over the whole soundboard and the modal analysis has shown that the soundboard was similar to a homogeneous plate in this frequency range. Whatever the wavenumbers are (depending on the boundary conditions, for example), the modal shapes can be approximated locally by the product of two sines: 
\begin{equation}
\Phi(x,y)=\sin{(2\pi k_x x+\varphi_x)}\,\sin{(2\pi k_y y +\varphi_y)}
\end{equation}

Since we do not want to take into account the details of the geometry, location, etc., we may as well replace the local dependency of any modal shape by a random distribution:
\begin{equation}
\Phi_\nu(x_A,y_A)=\sin{(2\pi \alpha +\varphi)}\,\sin{(2\pi \beta +\varphi)}
\end{equation}
where the random quantities $\alpha$ and $\beta$ are uniformly distributed in $[$0,1$]$. 

For this type of modal shape and in order to obtain the orthogonality in terms of the mass matrix, we consider that the modal mass of any mode is $m_\nu=M/4$. For this low-frequency range, $M$ is the mass of the whole soundboard (including ribs, bridges and the two fir bars) and almost equal to 9~kg for our upright piano.

The spacing between successive modal frequencies $f_{\nu+1}-f_{\nu}$ is considered as the reciprocal of the modal density, as modelled in \S\ref{sec:ModDens}. The first mode is slightly different from one piano to another; a typical value of 70 Hz was chosen here. In order to account for the non-regularity of the soundboard geometry, a random proportion of the inter-modal interval is added to each modal frequency:
\begin{equation}
f_{\nu+1}=f_{\nu}+\dfrac{1}{n(f_\nu)}[1+\alpha]
\end{equation}
where $\alpha$ is a random quantity uniformly distributed in $[$1/2,+1/2$]$.

According to the experimental results of modal analysis, we took 2\% as a uniform value for modal loss factors $\eta_\nu$.

\subsection{Limit of the low-frequency range}
\label{sec:LowFreqLimit}
%XXXKE pas simple à expliquer...
It was shown in \S~\ref{sec:ModDens} that the vibration regime of the soundboard changes at around 1.1 kHz, that is when $k_{x}=\pi/p$ ($p$: inter-rib spacing). In general, 1.1 kHz should thus be considered as the limit of the low-frequency regime. However, this global description must be considered more carefully when waves are generated locally (by a string, for example) and that a local singularity of the structure, namely the bridge, alters significantly the wave-guide dynamics. The impedance of the bridge -- a slightly curved bar, almost orthogonal to the direction of the ribs -- is much larger than that of the table (or the ribs) for frequencies higher than the kHz (almost five times greater at 4kHz). % XXXKE for frequencies higher than the XXX kHz. % n'est-ce pas le cas en général ? 
The dispersion curve of transverse waves in the bridge only, considered as a beam, is shown in Fig.~\ref{fig:disperschevalet}. It appears that bending in the $x$-direction that would correspond to $k_{x}=\pi/p$ cannot occur below 4.2 kHz\footnote{Surprisingly (or not\ldots) this frequency is the fundamental frequency of $C_8$, the highest piano note.}. Below this frequency, the bridge prevents the vibration from being localised in a single waveguide (or in three of them, with the same $k_x=\pi/p$); in effect, it makes the system {plate+bridge+ribs} more or less homogeneous up to $\approx4.2$~kHz. As a first approximation, we consider that the low-frequency regime for the mobility at the bridge extends up to 4.3 kHz.
\Figure{fig:disperschevalet}{Dispersion law of the flexural waves in the bridge.}
  {48}{  \put(2,1)  {  \includegraphics[width=78mm, height=47mm]{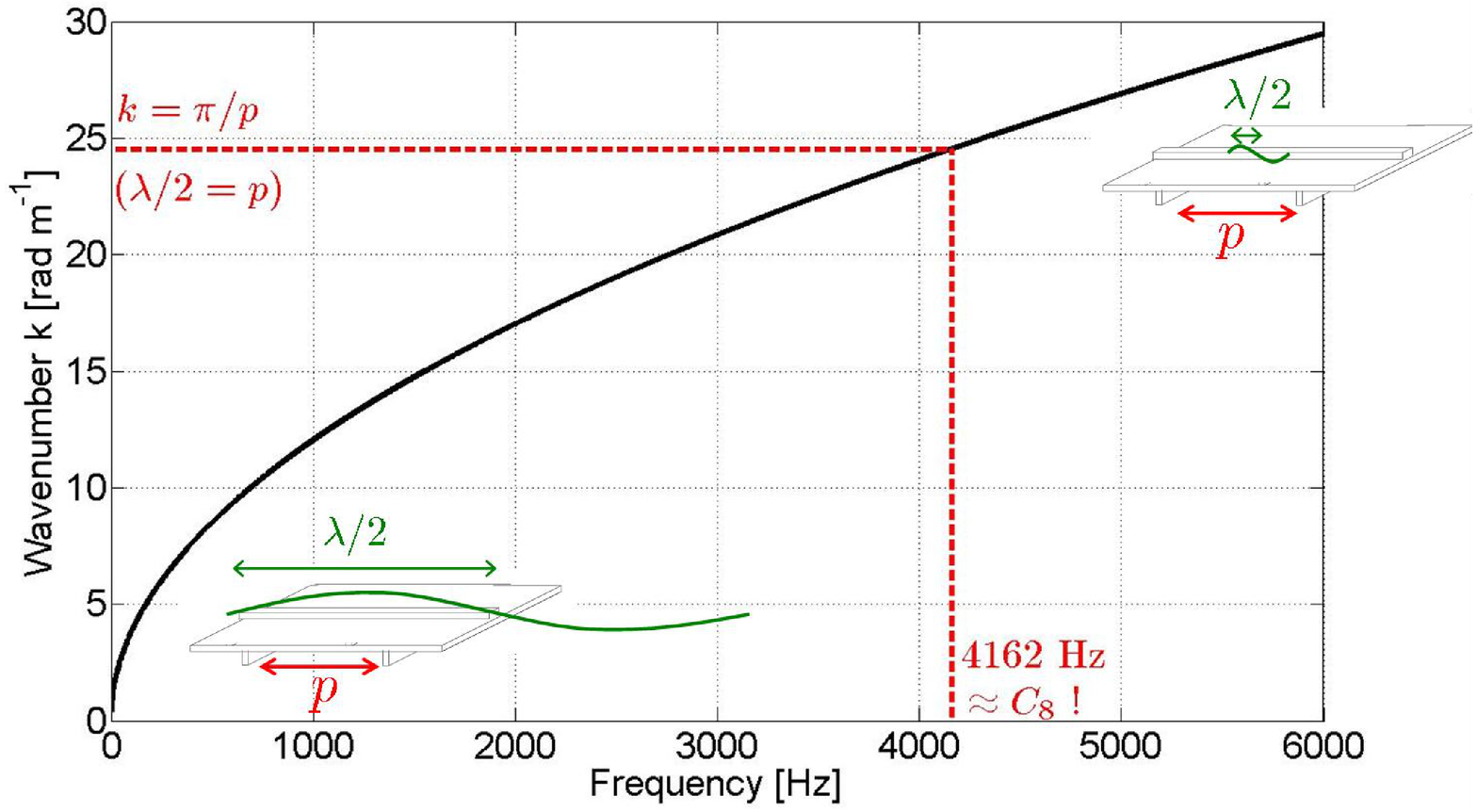}}}
  
\subsection{Mobility above 1.1 kHz far from the bridge}
\label{sec:FarBridge}
The approach is the same as in the low-frequency range. As explained above, modes above 1.1 kHz are localised in a region extending over approximately three inter-rib bands, with $k_{x}=\pi/p$. The reasoning on the modal shapes can thus be repeated \emph{verbatim}. The mass $M$ is considered to be three times the mass located between two ribs, \emph{i.e.} $\approx$0.7 kg overall. The modal masses are here also $m_\nu=M/4$. The modal frequencies are taken as explained in \S~\ref{sec:ModDens} for the case of the wave-guide. Although the acoustical radiation regime changes in that frequency region, we consider that the modal loss factors are the same as in the low-frequency region. This approximation needs to be reconsidered in a future research, when a global approach, similar to that performed on the vibrations, will have been done on the acoustical radiation of the soundboard.

We present in figure~\ref{fig:Y_1200} the real part of the synthetised mobility, according to Eq.~\ref{eq:drive_admitt_amort},  \emph{far from the bridge}. 
%The ed admittance tends towards the theoretical asymptote (XXX Skud), and the envelope given by the second calculus of Langley~\eqref{eq:G_resLang} coincides for the whole spectrum with the resonances and antiresonances of the syntesized admittance.
The magnitude of the synthesised impedance is presented in Fig.~\ref{fig:Z_1200} and compares very well with published measurements by Giordano~\cite{GIO1998}. The fall of impedance measured around 1~kHz is well predicted by our model. %XXXKE (garder cette phrase) ?
%We add also on the graph presenting the measurements of Giordano the calculated mean value of the impedance (from Skudrzyk). The curves match well... decrease of impedance...
\Figure{fig:Y_1200}{Real part of the mobility \textit{far from the bridge}. -----~: synthesised mobility. {\color[rgb]{0,0,1}--~--~--}~: mean value (see~\S\ref{sec:mean}). {\color[rgb]{0,0,1}------}~: envelopes (see~\S\ref{sec:envelope}).}
  {48}{  \put(2,1)  {\includegraphics[width=78mm, height=47mm]{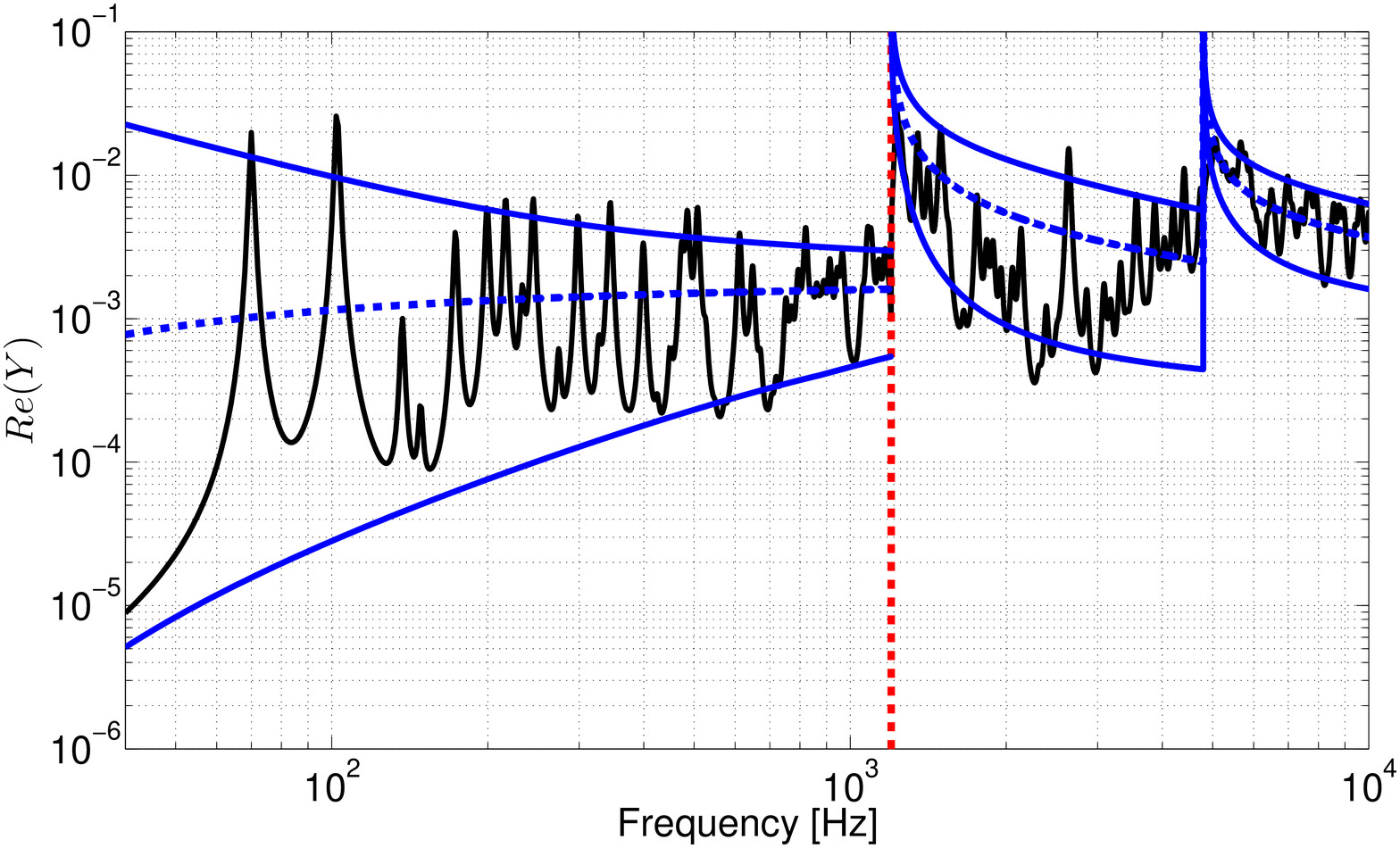}}}
\Figure{fig:Z_1200}{Magnitude of the impedance \emph{far from the bridge}. Top: synthesised values. Bottom: measurements published by Giordano~\cite{GIO1998}. {\color[rgb]{0,0,1}--~--~--}: mean value according to Skudrzyk (see~\S\ref{sec:mean}).}
  {94}{  \put(4,48)  {\includegraphics[width=78mm, height=47mm]{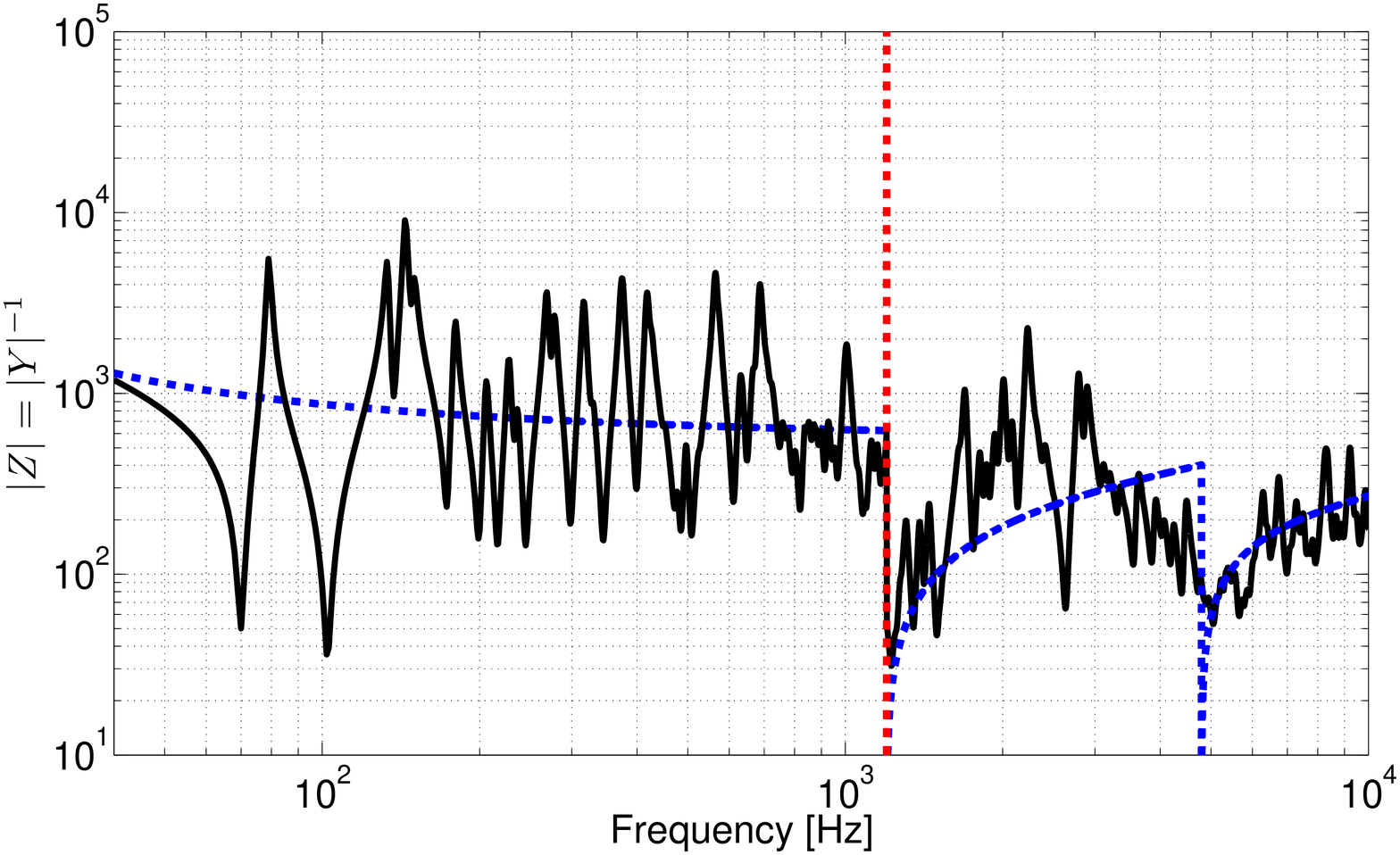}} \put(2,1) {\includegraphics[width=75mm]{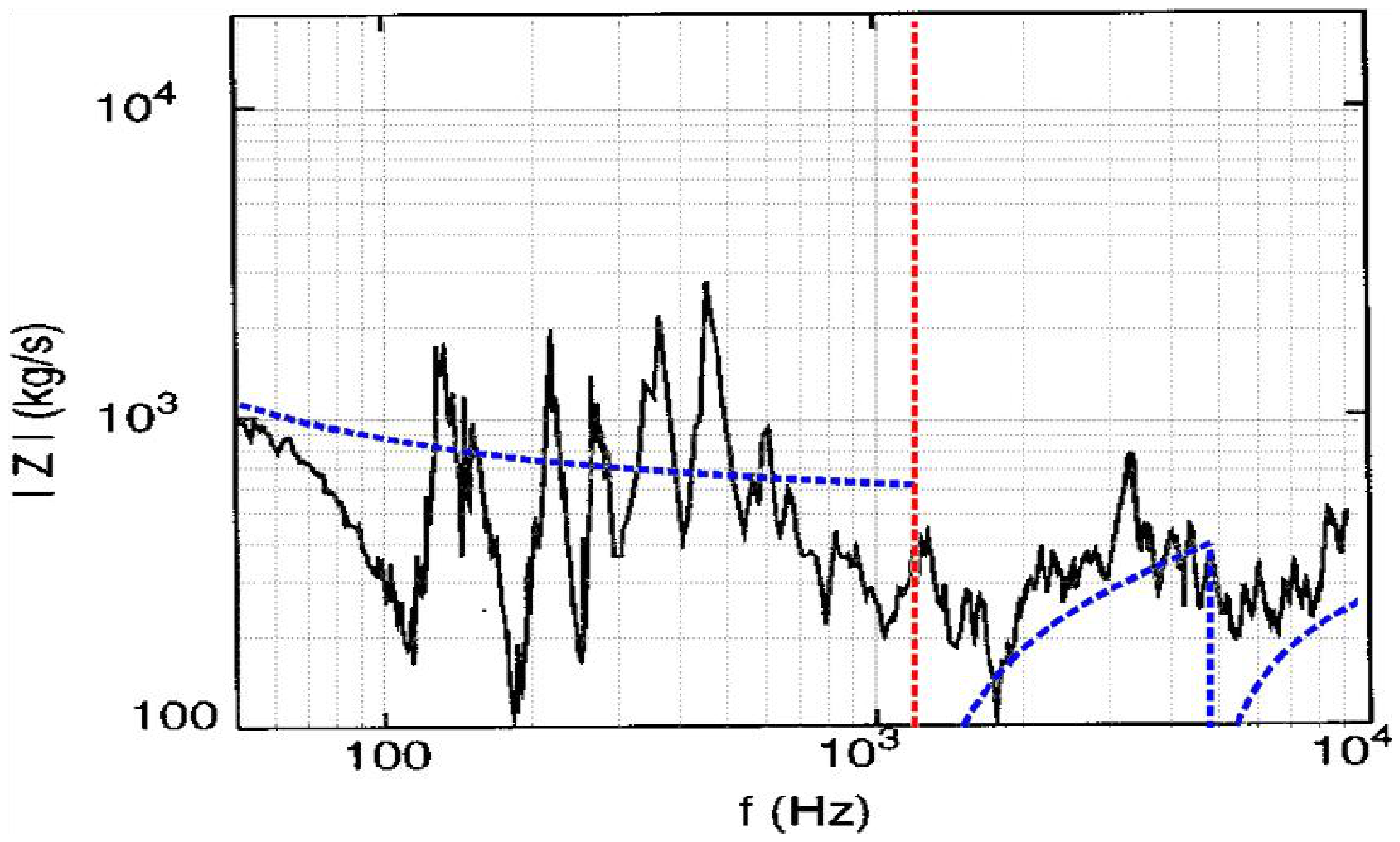}}}

\subsection{Mobility above 1.1 kHz at the bridge}
As explained in \S~\ref{sec:LowFreqLimit}, we considered that the bridge extends the frequency range where the soundboard can be considered as homogeneous, namely up to 4.2 kHz. Beyond that limit, we adopt for the mobility at the bridge the same approach as for the mobility in a wave-guide, far from the bridge, as described in \S~\ref{sec:FarBridge}. There is clearly here a lack of selfconsistency which, hopefully, will be resolved by future research.

The synthesised mobility far from the bridge is given in figure~\ref{fig:Y_4200}) and the magnitude of the synthesised impedance in Fig.~\ref{fig:Z_4200}. The mean value of the synthesised impedance between 100 and 1000~Hz is approximately 800~kg~s$^{-1}$. This value is consistent with the measurements at the bridge published by Wogram~\cite{WOG1980} or Giordano~\cite{GIO1998}: these authors measured a mean impedance for typical upright piano of about $10^3$~kg~s$^{-1}$. Moreover, the fluctuations of the mobility for those frequencies are $\pm10$-$15$~dB, which is also consistent with measurements published by Conklin~\cite{CON1996_2} for example.
Nevertheless a discrepency on the average value of the impedance (particularly visible around 1-3kHz, on figure~\ref{fig:Z_4200}) exists. Our model must be improved in order to take into account more properly the influence of the bridge. %XXXKE the average value must be increased.
\Figure{fig:Y_4200}{Real part of the mobility \textit{at the bridge}. -----~: synthesised mobility. {\color[rgb]{0,0,1}--~--~--}~: mean value (see~\S\ref{sec:mean}). {\color[rgb]{0,0,1}------}~: envelopes (see~\S\ref{sec:envelope}).}
  {48}{  \put(2,1)  {  \includegraphics[width=78mm, height=47mm]{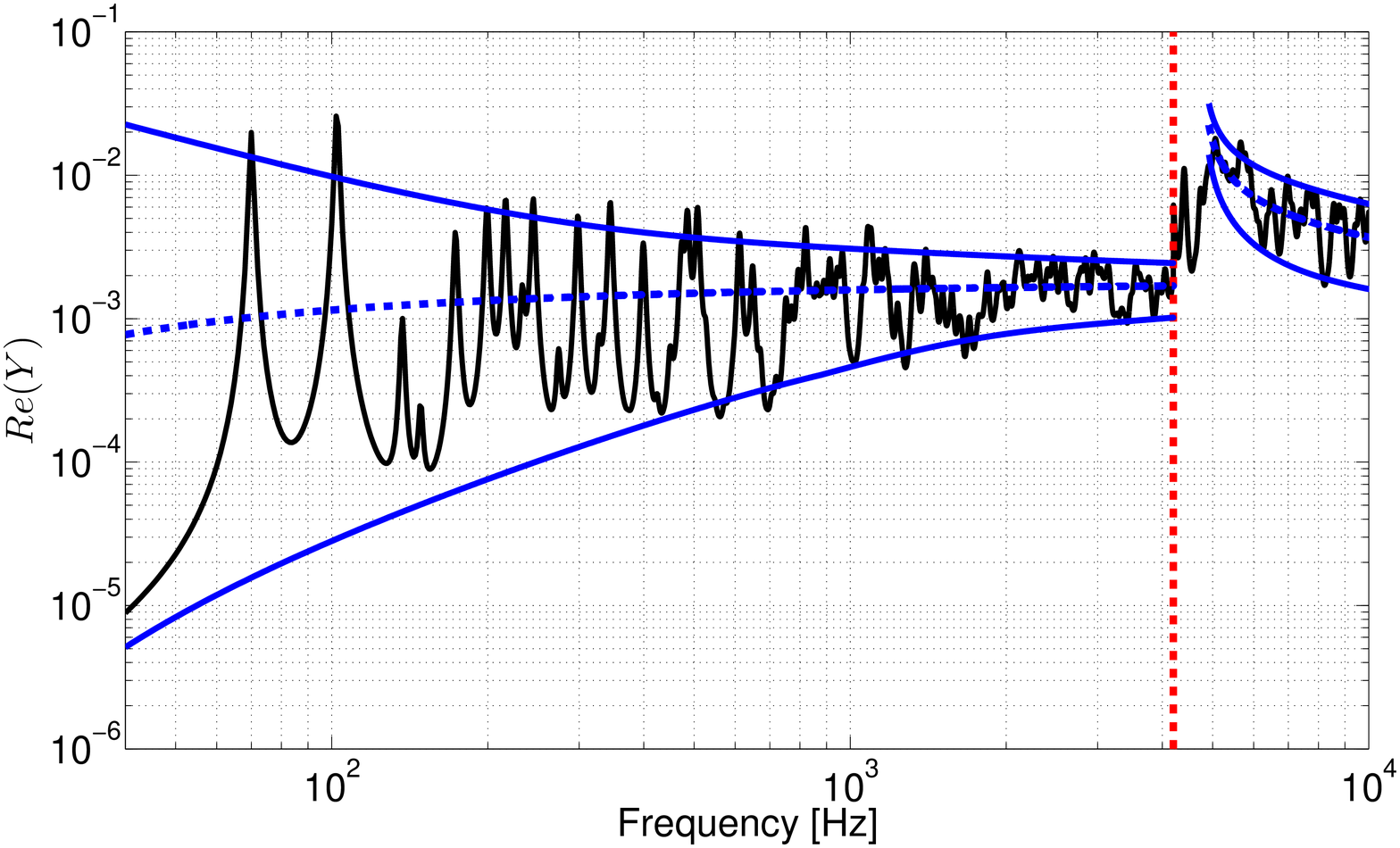}}}
 \Figure{fig:Z_4200}{Magnitude of the impedance \emph{at the bridge}. Top: synthesised values. Bottom: measurements published by Giordano~\cite{GIO1998}. {\color[rgb]{0,0,1}--~--~--}: mean value according to Skudrzyk (see~\S\ref{sec:mean}).}{94}{  \put(4,48)  {\includegraphics[width=78mm, height=47mm]{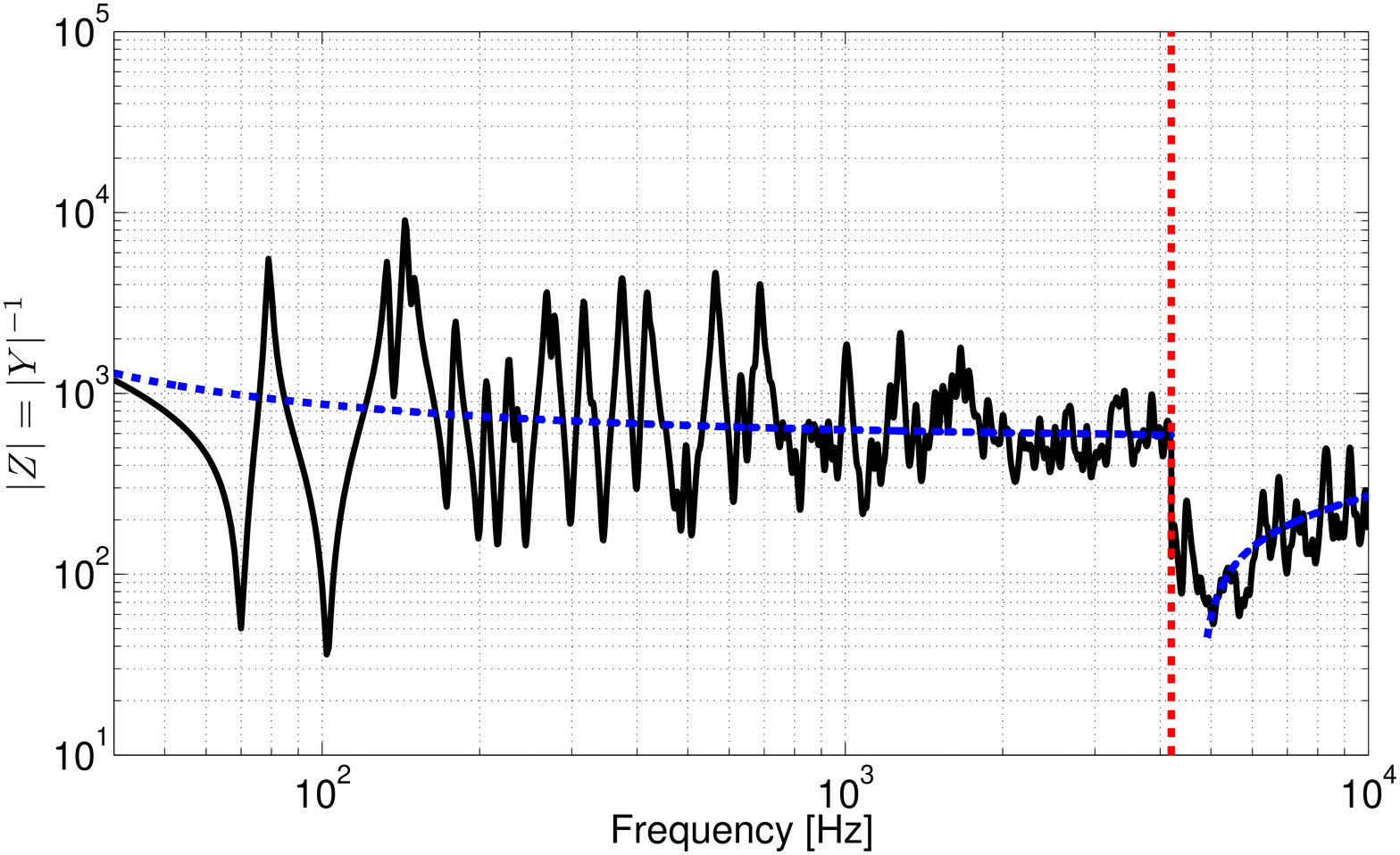}} \put(2,1) {\includegraphics[width=75mm]{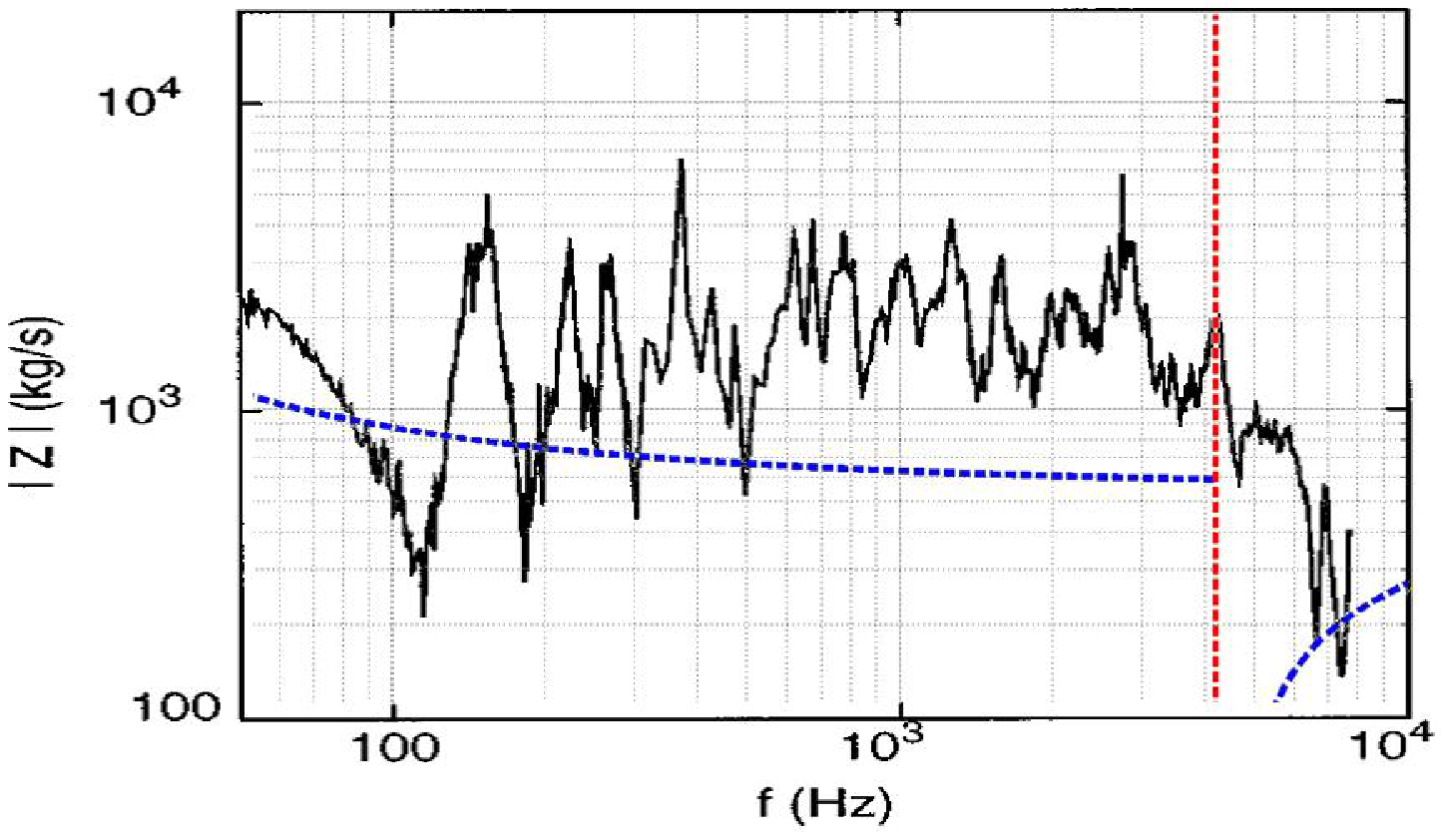}}}

\section{Comparison with global approaches}
\label{sec:global}
\subsection{Skudrzyk mean-value theorem}\label{sec:mean}
Skudrzyk's \emph{mean-value method} (proposed in \cite{SKU1958}--\cite{SKU1968} and theorised in its final form in \cite{SKU1980}) predicts the mean value and the asymptotic value of the driving point admittance of a weakly dissipative vibrating structure. This approach is not frequency limited.

The principal results obtained by Skudrzyk are recalled here. Skudrzyk's idea consists in replacing the discrete summation in equation~\ref{eq:drive_admitt_amort} by a continous integral. After simplification of the denominator in the hypothesis of small damping, the transformation of  equation \ref{eq:drive_admitt_amort} takes the form:
%\begin{equation}\label{eq:Y_C}
%Y_\text{A}(\omega)\underset{\omega\rightarrow+\infty}{\rightarrow}Y_C=\int_0^{+\infty}{\cfrac{i\omega\Phi_\nu^2(x_\text{A},y_\text{A})\,\text{d}\omega_\nu}{ m_\nu\epsilon_\nu\,(\omega_\nu^2+i\eta_\nu\omega_\nu\omega-\omega^2)}}
%\end{equation}
%where $\epsilon_\nu=\cfrac{\text{d}\omega_\nu}{\text{d}\nu}=2\pi\,\Delta\!f_\nu=\cfrac{2\pi}{n(f_\nu)}$ is the average modal spacing (written here for pulsations and corresponding to the inverse of the modal density $n(\omega_\nu)$). The writing of the denominator of $Y_\text{C}$ can be simplified in the hypothesis of small damping (see \cite{SKU1958} or \cite{CRE2005}) and the equation \eqref{eq:Y_C} takes the form:
\begin{equation}\label{eq:Y_C2}
Y_\text{C}=\int_0^{+\infty}{\cfrac{i\omega\Phi_\nu^2(x_\text{A},y_\text{A})}{m_\nu\,\epsilon_\nu\,(\bar{\omega}_\nu^2-\omega^2)}}\,\text{d}\omega_\nu=G_\text{C}+i\,B_\text{C}
\end{equation}
where $\bar{\omega}_\nu^2=\omega_\nu^2(1+i\eta_\nu)$, $G_\text{C}=\Re(Y_\text{C})$, $B_\text{C}=\Im(Y_\text{C})$, and $\epsilon_\nu=\cfrac{\text{d}\omega_\nu}{\text{d}\nu}=2\pi\,\Delta\!f_\nu=\cfrac{2\pi}{n(f_\nu)}$ is the average modal spacing, written here for angular frequencies. Finally, by use of the residue theorem, the real part of the driving point admittance is given by:
\begin{equation}\label{eq:Skud_admittmean}
\Re(Y_A(\omega))\underset{+\infty}{\rightarrow}G_\text{C}=\cfrac{\pi}{2\,\epsilon_\nu\,M}=\cfrac{n(f)}{4M}
\end{equation}
In this frequency domain, the real part of the admittance depends only on the modal density and the mass of the structure. For a thin plate, the imaginary part $B_\text{C}$ vanishes at high frequency \cite{SKU1980}:
\begin{equation}\label{eq:admitt_HF}
Y_A(\omega)\underset{+\infty}{\rightarrow} G_\text{C}=\cfrac{1}{4h^2}\sqrt{\cfrac{3(1-\nu_{xy}^2)}{E\,\rho}}
\end{equation}
written here in the isotropic case. $G_\text{C}$ is equivalent to the driving point admittance of the infinite plate~\cite{CRE2005}. It depends neither on the frequency, nor on the surface but only on the thickness $h$ and on the elastic constants of plate: the Young's modulus $E$, the Poisson's ratio $\nu_{xy}$, and the density~$\rho$.
By extrapolating towards the low frequencies, Skudrzyk's theory predicts the mean value of the admittance: $G_\text{C}=\Re{(Y_\text{C})}$ is the geometric mean of the values at resonances $G_\text{res}$ and antiresonances $G_\text{ares}$.

%\subsection{Envelope}\label{sec:envelope}
%Skudrzyk gives an approached expression of the envelope of the resonances and antiresonances. Under the assumptions of well-separated peaks and equal modal masses (peaks of the impulses responses of equal amplitudes), a single-degree of freedom damped oscillator~$\nu$ has an amplitude at resonance $f_\nu$ of:
%\begin{eqnarray}\label{eq:Skud_Res}
%G_{\text{res}}\approx\cfrac{1}{\eta\omega_\nu\,M}=\cfrac{n(f_\nu)}{4M}\:\cfrac{2}{\pi\mu(f_\nu)}=G_\text{C}\:\beta(f_\nu)
%\end{eqnarray}
%with $\beta(f)=\cfrac{2}{\pi\mu(f)}=\cfrac{2}{\pi n(f)\eta f}$ and where the indicator $\mu(f)=n(f) \eta f$ is the modal overlap factor defined as the ratio between the half-power modal bandwidth and the average modal spacing. Generally, $\mu$ increases with frequency, and thus the amplitude of resonances decreases. In the theory of Skudrzyk, $G_\text{C}$ is the geometric mean value of the admittance (for all the frequencies) $G_\text{C}=(G_\text{res}G_\text{ares})^{1/2}$. This yields directly the amplitude of antiresonances:
%\begin{equation}\label{eq:Skud_aRes}
%G_{\text{ares}}\approx\cfrac{G_\text{C}}{\beta(f_\nu)}=\cfrac{n(f_\nu)}{4M}\:\cfrac{\pi\mu(f_\nu)}{2}
%\end{equation}

\subsection{Langley's envelopes calculations}\label{sec:envelope}
Langley~\cite{LAN1994} evaluates analytically the envelopes of the analytical summation given in equation~\ref{eq:drive_admitt_amort}.
Bidimensional structures, such as plates, can present repeated resonances, degeneracy and thus irregular modal spacing. %This severely degrades the accuracy of the admittance envelope given by \eqref{eq:G_resLang} and \eqref{eq:G_aresLang}.
Langley introduces semi-empirical modifications in order to take into account these irregularities. His approach is derived from the one of repartition of the resonances in room acoustics. %\cite{BOL1946}-\cite{BOL1947} or \cite{SEP1965}.
Under the assumption that the modal spacing conforms to the \emph{Poisson's law}, the amplitudes of resonance frequencies of a bi-dimensional rectangular structure are given by (\cite{LAN1994}):
\begin{eqnarray}
G_{\text{res}}\approx G_\text{C}\:(1+\frac1{\sqrt{\mu_2}})\:\coth{[(1+\frac1{\sqrt{\mu_2}})\:\frac{\pi\mu_2}{2}]}
\label{eq:G_resLang2}
\end{eqnarray}
where the modal overlap factor $\mu(f)=n(f) \eta f$ (defined as the ratio between the half-power modal bandwidth and the average modal spacing) is modified in $\mu_2=[1-(L_1 L_2)^{-1}]\,\mu\,$ in order to take into account the repeated frequencies. $\mu_2$ depends on the natural numbers $L_1$ and $L_2$ related to the aspect ratio of the rectangular structure by $L_2/L_1=L_y/L_x$. 
%\footnote{Par exemple, la pulsation propre $\omega_{mn}$ d'une plaque rectangulaire isotrope de facteur d'aspect ${L_y}/{L_x}=L_2/L_1$ sera \emph{répétée} si $L_2\,n/L_1$ et $L_1 m/L_2$ sont entiers.} (supposée rectangulaire) par $L_2/L_1=L_y/L_x$.
Similarly, Langley gives the amplitude of antiresonances $G_\text{ares}$.

\subsection{Comparisons} %XXXKE intégrer éventuellement cette sous-section aux deux précédentes ?
The mean values $G_C$ and envelopes $G_\text{res}$ and $G_\text{ares}$ are represented in figure~\ref{fig:Y_1200}-\ref{fig:Z_4200}. The synthesised quantities displays the right properties in terms of mean value, asymptot, and envelopes.

\section{Conclusion}
The approach presented in this communication avoids the detailed description of the soundboard, based on a very large number of parameters. It can be used to predict global changes of the driving point mobility, and possibly of the sound radiation in the treble range, resulting from structural modifications. Synthesised impedances match well with published measurements. Our model must be improved in order to take into account more properly the influence of the bridge. %XXXKE problème ouvert.
%XXKE cf notre article biblio et la remarque de Conklin : 
Let's conclude with a remark by Conklin~\cite{CON1996_2} concerning the effect of bridges on the tone production. \emph{The design of the soundboard bridges affects profoundly the tone of a piano. %The bridges and soundboard together determine the load presented to the strings. 
In coupling the strings to the soundboard, the bridges functions as impedance transformers presenting a higher impedance to the strings than would exist in the case of direct coupling. \emph{[...]} If the strings were terminated directly on the soundboard, the result would be a louder-than-normal but relatively unpleasant tone of comparatively short duration. By adjusting the design of the bridges, the designer of a piano can change the loudness, the duration, and the quality of the tone, within a certain range, in order to suit the intended use of the instrument.} Some more research is clearly needed in order to put some numbers onto these assertions.

~%\section{CONCLUSIONS}

%This template can be found on the conference website. This template can be found on the conference website.  

% Acknowledgement (activate if applicable)
%\Acknowledgement{This project has been funded by the research council of country 1.}

\end{document}